\documentclass[aps,prl,reprint,floatfix]{revtex4-2}
\usepackage{amsmath,amssymb,bm,graphicx,url}
\usepackage{xcolor}
\definecolor{linkblue}{rgb}{0.1,0.25,0.6}
\usepackage[colorlinks=true,linkcolor=linkblue,citecolor=linkblue,urlcolor=linkblue]{hyperref}
\DeclareMathOperator{\Cov}{Cov}
\DeclareMathOperator{\Var}{Var}
\begin{document}
\title{Speed--response trade-off in non-genetic adaptation of a continuous phenotype}
\author{Henry Mattingly}
\affiliation{Center for Computational Biology, Flatiron Institute, New York, NY 10010, USA}
\begin{abstract}
Non-genetic inheritance lets populations adapt without mutations. For a continuous trait, we decompose the phenotype distribution and selection pressure into eigenmodes of the offspring-generating process, each with its own relaxation time. We show that the response of each mode to selection trades off with its relaxation rate. For cells selected on their division rate in an environment that reverses periodically, we derive the optimal inheritance that maximizes long-term population growth. These results relate the strength of inheritance to the speed and extent of adaptation.
\end{abstract}
\maketitle

Genetically identical individuals differ, and offspring often inherit traits that are not fixed by their genes, such as gene expression levels~\cite{Sigal2006,Hormoz2015,Shaffer2020}, cell size~\cite{Amir2014,TaheriAraghi2015}, growth rate~\cite{Kiviet2014,Sandler2015}, or swimming behavior~\cite{Pleska2021}. Models of genetic evolution usually neglect this heritable variation within a genotype~\cite{Crow1970,Ewens2004,Falconer1996,Lynch1998}. Yet selection can act on it and change the phenotype distribution without mutations~\cite{Jablonka1995,Lachmann1996,Bonduriansky2009,Bonduriansky2018,Day2016,Hashimoto2016,Lin2017,Nozoe2017,Lin2020,Levien2020,Levien2021}. For example, collectively migrating populations of \textit{Escherichia coli} adapt their distribution of swimming and sensing phenotypes to the environment~\cite{Mattingly2022,Vo2025}, and some cancer cells survive drug treatment in rare, partly heritable states~\cite{Sharma2010,Shaffer2017,Marine2020,Emert2021,Goyal2023}. Because genes are selected only through the phenotypes they produce, this non-genetic selection can modify genetic evolution~\cite{Day2011,Bonduriansky2012,Shih2018,Carja2019}. A theory of non-genetic selection is therefore needed to understand these effects.

Theory has described non-genetic selection in two separate settings. In constant environments, it relates parent--offspring correlations of continuous traits, measured in cell lineages~\cite{Kiviet2014,Hormoz2015,Sandler2015,Pleska2021}, to the growth of asynchronously dividing populations~\cite{Hashimoto2016,Lin2017,Lin2020,Levien2020}. In changing environments, models instead describe populations that switch stochastically between two phenotypes, as in bacterial persistence~\cite{Balaban2004,Kussell2005b} and bet hedging~\cite{Cohen1966,Thattai2004,Kussell2005,Wolf2005,Acar2008,Beaumont2009,Xue2017}, or strategies with memory or plasticity~\cite{Rivoire2011,Lambert2014,Rivoire2014,Xue2016,Skanata2016,Mayer2017,Xue2018,Xue2019,George2023}. These models find that switching rates should track the rates of environmental change~\cite{Kussell2005} and that an intermediate strength of inheritance can be optimal~\cite{Lachmann1996,Carja2017}. Many of them also count offspring per synchronized generation~\cite{Cohen1966,Lachmann1996,Rivoire2014,Carja2017}.

Combining continuous traits and asynchronous division with changing environments raises new questions. The distribution of a continuous trait has many features, such as the mean and the variance, and inheritance erases each on its own timescale. Which of these timescales sets how far and how fast a population adapts, and what amount of inheritance optimizes population growth when the environment keeps changing?

Here we study how the phenotype distribution of an isogenic population of cells responds to selection when a continuous, heritable phenotype sets their division rate. We decompose this distribution and selection into eigenmodes of the offspring-generating process and derive an exact equation for each. Each mode faces a trade-off: a mode that relaxes slowly builds a large response to selection but keeps it after the environment changes, whereas a mode that relaxes quickly follows changes but barely responds. When selection reverses periodically, this trade-off sets the inheritance that maximizes long-term growth: the selected mode should relax in about a third of each epoch, plus one division time. These results are a step toward understanding how non-genetic inheritance shapes the ecology and evolution of genotypes.

We consider one genotype, so all phenotypic variation is non-genetic. Let the offspring distribution $\mathcal P(x'|x)$ be the probability density, normalized over $x'$, that an offspring of a parent with phenotype $x$ has phenotype $x'$. We assume a unique stationary distribution $p^{(0)}$, which parents pass on unchanged to their offspring,
\begin{equation}
 \int\mathcal P(x|y)\,p^{(0)}(y)\,dy=p^{(0)}(x).
 \label{eq:p0}
\end{equation}
We measure the phenotype from its stationary mean in units of its stationary standard deviation, so that $x$ has mean zero and variance one under $p^{(0)}$. We assume that inheritance is parent--offspring symmetric: the stationary joint distribution of parent and offspring phenotypes is unchanged by exchanging the two, $p^{(0)}(y)\,\mathcal P(x|y)=p^{(0)}(x)\,\mathcal P(y|x)$.

\begin{figure}[tb]
\includegraphics[width=\columnwidth]{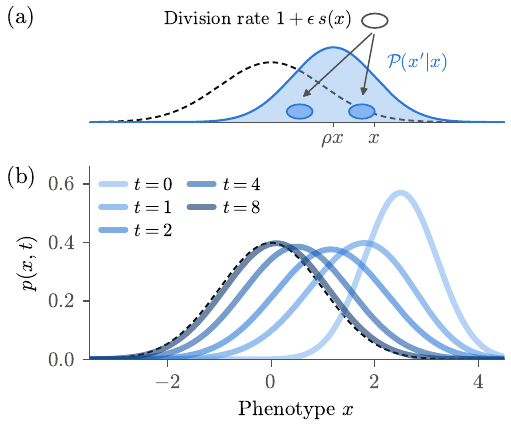}
\caption{Inheritance and relaxation. (a) A parent with phenotype $x$ divides at rate $1+\epsilon\,s(x)$ into two daughters whose phenotypes are drawn from $\mathcal P(x'|x)$, here a normal distribution centered at $\rho\,x$. (b) Without selection, an initial distribution relaxes to $p^{(0)}$, for Gaussian inheritance with $\rho=0.8$. The mean relaxes on $\tau_1=2.5$ and the second moment on $\tau_2\approx1.4$ division times. Thick lines: exact expected-density equation; dashed lines: $p^{(0)}$.}
\label{fig:model}
\end{figure}

Let $n(x,t)$ be the expected population density. The population size and the phenotype distribution are
\begin{equation}
 N(t)\equiv\int n(x,t)\,dx,\qquad p(x,t)\equiv\frac{n(x,t)}{N(t)}.
 \label{eq:Np}
\end{equation}
In binary fission, a dividing parent is replaced by two daughters, each of whose phenotypes is drawn from $\mathcal P$ [Fig.~\ref{fig:model}(a)]. The Supplemental Material (SM) treats budding, in which the parent remains. We measure time in units of the mean division time without selection, and let selection change the division rate of phenotype $x$ to $1+\epsilon\,s(x)$, where $\epsilon$ sets the strength of selection (the SM treats selection on the death rate). The expected density then obeys
\begin{equation}
 \dot n(x,t)=\int[2\,\mathcal P(x|y)-\delta(x-y)][1+\epsilon\,s(y)]\,n(y,t)\,dy.
 \label{eq:fission}
\end{equation}
The delta function removes the dividing parent.

Each division adds one individual net, so the population growth rate equals the mean division rate,
\begin{equation}
 \Lambda(t)\equiv\frac{\dot N(t)}{N(t)}=1+\epsilon\int s(x)\,p(x,t)\,dx.
 \label{eq:Lambda}
\end{equation}
Differentiating $p=n/N$ then gives the dynamics of the phenotype distribution,
\begin{multline}
 \dot p(x,t)=\int[2\,\mathcal P(x|y)-\delta(x-y)][1+\epsilon\,s(y)]\,p(y,t)\,dy\\
 -\Lambda(t)\,p(x,t).
 \label{eq:pdot}
\end{multline}
We center the selection profile, \mbox{$\int p^{(0)}(x)\,s(x)\,dx=0$}, because a constant part of $s$ only rescales time and $\epsilon$.

We describe the phenotype distribution through inheritance modes $\psi_i$, functions of the phenotype whose expected value in an offspring is proportional to their value in the parent,
\begin{equation}
 \int\mathcal P(x'|x)\,\psi_i(x')\,dx'=\lambda_i\,\psi_i(x).
 \label{eq:left}
\end{equation}
The constant function has eigenvalue one, and all other eigenvalues are smaller. For symmetric inheritance, the eigenvalues are real, and the modes can be chosen to be orthonormal with weight $p^{(0)}$,
\begin{equation}
 \int p^{(0)}(x)\,\psi_i(x)\,\psi_j(x)\,dx=\delta_{ij}.
 \label{eq:ortho}
\end{equation}

Selection moves the phenotype distribution away from $p^{(0)}$. We express this relative change in terms of the inheritance modes,
\begin{equation}
 p(x,t)=p^{(0)}(x)\,\Big[1+\epsilon\sum_{i\geq1}\chi_i(t)\,\psi_i(x)\Big],
 \label{eq:expansion}
\end{equation}
which defines the mode amplitudes $\chi_i$. The constant mode contributes only the one, because $p$ is normalized. Multiplying Eq.~\eqref{eq:expansion} by $\psi_i(x)$, integrating, and using Eq.~\eqref{eq:ortho} shows that each amplitude is the average of its mode over the current population,
\begin{equation}
 \chi_i(t)=\langle\psi_i\rangle/\epsilon,\qquad
 \langle f\rangle\equiv\int f(x)\,p(x,t)\,dx.
 \label{eq:chii}
\end{equation}
Likewise, we expand the centered selection profile in the modes, with selection coefficients $s_i$,
\begin{equation}
 s(x)=\sum_{i\geq1}s_i\,\psi_i(x),\quad
 s_i\equiv\int p^{(0)}(x)\,\psi_i(x)\,s(x)\,dx.
 \label{eq:si}
\end{equation}
Inserting both expansions into Eq.~\eqref{eq:Lambda} gives the growth rate exactly as a sum over modes,
\begin{equation}
 \Lambda(t)=1+\epsilon^2\sum_{i\geq1}s_i\,\chi_i(t).
 \label{eq:Lambdamodes}
\end{equation}

Multiplying Eq.~\eqref{eq:pdot} by $\psi_i(x)$, integrating over $x$, and using Eq.~\eqref{eq:left} gives an exact equation for each amplitude (see SM),
\begin{gather}
 \tau_i\,\dot\chi_i(t)=(\tau_i-1)\Cov(\psi_i,s)-\Lambda(t)\,\chi_i(t),
 \label{eq:modeexact}\\
 \tau_i\equiv\frac{1}{2\,(1-\lambda_i)},
 \label{eq:tau}
\end{gather}
where $\Cov$ and $\Var$ denote the covariance and variance over the current population. Each mode relaxes on the timescale $\tau_i$, where the factor two counts the two inheritance events per division, and it is forced by its covariance with the selection profile in the current population, as in Price's equation~\cite{Price1970,Day2011}. Selection also couples the modes, because both $\Lambda$ and the covariance depend on the entire current population.

In a constant environment, the population approaches a steady state, which we mark by a superscript $\infty$. Setting the time derivative in Eq.~\eqref{eq:modeexact} to zero gives the steady response of each mode,
\begin{equation}
 \chi_i^\infty=\frac{(\tau_i-1)\Cov^\infty(\psi_i,s)}{\Lambda^\infty}.
 \label{eq:steadyexact}
\end{equation}
This relation is exact, for any offspring distribution and selection profile.

For weak selection, $\Lambda$ is close to one and the population stays close to $p^{(0)}$, so the covariance in Eq.~\eqref{eq:modeexact} is close to $s_i$. The mode equation then becomes
\begin{equation}
 \tau_i\,\dot \chi_i(t)=(\tau_i-1)\,s_i-\chi_i(t),
 \label{eq:mode}
\end{equation}
so each amplitude relaxes on the timescale $\tau_i$, driven only by its own selection coefficient, toward the steady response
\begin{equation}
 \chi_i^\infty=(\tau_i-1)\,s_i.
 \label{eq:chiinf}
\end{equation}

This is the speed--response trade-off: at a fixed selection coefficient, the more slowly a mode relaxes, the more strongly it responds. The same trade-off holds for budding and for selection on the death rate (see SM). The response has the sign of $s_i$ only for modes that relax more slowly than one division time, or, by Eq.~\eqref{eq:tau}, $\lambda_i > 1/2$, a threshold similar to one found for the effect of single-cell growth-rate variability on population growth~\cite{Lin2020}. Faster modes respond against selection: fitter cells divide more often, but each division replaces the parent by offspring that keep, on average, less than half of its excess division rate, $\epsilon\,s(x)$. By Eq.~\eqref{eq:Lambdamodes}, such modes lower the growth rate in a constant environment.

\begin{figure}[tb]
\includegraphics[width=\columnwidth]{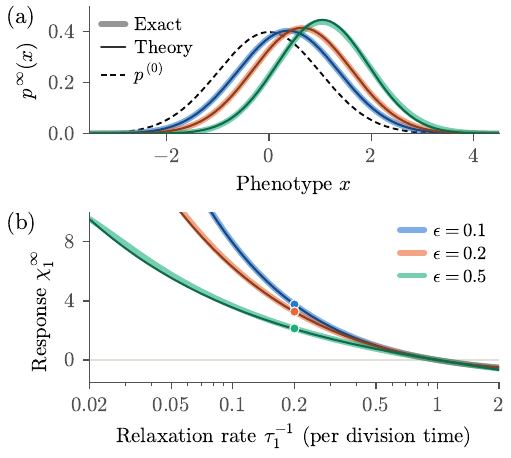}
\caption{Steady states under directional selection, $s(x)=x$, for Gaussian inheritance. (a) Phenotype distributions for $\rho=0.9$, so that $\tau_1=5$, and increasing selection strength $\epsilon$. (b) Steady response of the first mode against its relaxation rate, as $\rho$ varies from $0$ to $0.99$. Dots mark the distributions in (a). Thick lines: exact expected-density equation; thin lines: theory, Eqs.~\eqref{eq:var} and \eqref{eq:closed}, drawn in (a) as normal distributions with the predicted mean and variance; dashed line: $p^{(0)}$.}
\label{fig:steady}
\end{figure}

Beyond weak selection, the covariances depend on the shape of the offspring distribution. For a trait set by many small, partly inherited contributions, the joint distribution of parent and offspring phenotypes is close to Gaussian, as in quantitative genetics~\cite{Falconer1996}. We take the offspring phenotype to regress toward the stationary mean with parent--offspring correlation $\rho$, which lies between $-1$ and one,
\begin{equation}
 \mathcal P(x'|x)=\mathcal N\big(\rho\,x,\;1-\rho^2\big),\qquad
 p^{(0)}(x)=\mathcal N(0,\;1),
 \label{eq:gaussian}
\end{equation}
where $\mathcal N$ denotes a normal density with the given mean and variance. The inheritance modes are the normalized Hermite polynomials, the first two of which are
\begin{equation}
 \psi_1(x)=x,\qquad \psi_2(x)=\frac{x^2-1}{\sqrt2},
 \label{eq:hermite}
\end{equation}
with eigenvalues $\lambda_i=\rho^i$~\cite{Mehler1866}. Since $x^2=1+\sqrt2\,\psi_2(x)$, Eq.~\eqref{eq:chii} relates the first two mode amplitudes to the mean and the variance of the phenotype,
\begin{equation}
 \langle x\rangle=\epsilon\,\chi_1,\qquad
 \Var(x)=1+\sqrt2\,\epsilon\,\chi_2-\epsilon^2\,\chi_1^2,
 \label{eq:moments}
\end{equation}
and higher amplitudes to higher moments. Figure~\ref{fig:model}(b) shows an initial distribution relaxing to $p^{(0)}$ without selection, its mean on the timescale $\tau_1$ and its second moment on the shorter $\tau_2$. Here and below, we test the theory against the exact expected-density equation, solved numerically (see SM).

For positive $\rho$, the first mode, which shifts the mean, is the slowest, so by Eq.~\eqref{eq:chiinf} it responds most strongly to selection. We therefore consider directional selection, which acts on this mode alone, so that Eq.~\eqref{eq:Lambdamodes} has a single term,
\begin{equation}
 s(x)=x=\psi_1(x),\qquad
 \Lambda(t)=1+\epsilon^2\,\chi_1(t).
 \label{eq:smean}
\end{equation}
The covariance of the first mode with $s$ is then the variance of $x$, so Eq.~\eqref{eq:steadyexact} relates the response exactly to the variance,
\begin{equation}
 \chi_1^\infty=\frac{(\tau_1-1)\Var^\infty(x)}{\Lambda^\infty}.
 \label{eq:steady1}
\end{equation}
For modes slower than one division, strong selection thus reduces the weak-selection response, Eq.~\eqref{eq:chiinf}, in two ways. The population divides faster, so inheritance redraws phenotypes toward $p^{(0)}$ more often, which divides the response by $\Lambda^\infty>1$. And selection narrows the distribution, leaving less variation to act on, which multiplies the response by $\Var^\infty(x)<1$.

For the second mode, the covariance is $\sqrt2\,\epsilon\,\chi_1\Var(x)$ plus a term in the third cumulant of the distribution, which is small (see SM). Neglecting it, the steady state of the second mode and Eq.~\eqref{eq:moments} give the variance (see SM),
\begin{equation}
 \Var^\infty(x)=\frac{\Lambda^\infty}{1+\alpha\,(\Lambda^\infty-1)},\qquad
 \alpha\equiv\tau_1-2\,\tau_2+2.
 \label{eq:var}
\end{equation}
Here, $\alpha = 2-1/[2(1+\rho)]$, which exceeds one for $\rho>-1/2$, so the population narrows when selection raises the growth rate. The steady distributions also stay close to normal [Fig.~\ref{fig:steady}(a)]. Equations~\eqref{eq:steady1} and \eqref{eq:var} use only that $x$ and $x^2-1$ are inheritance modes. This holds whenever the offspring phenotype is $\rho$ times the parent's plus independent noise, even if inheritance is then not parent--offspring symmetric. Normality enters only through the neglected third cumulant, which is small for normal noise but generally not for other noise (see SM).

Inserting Eq.~\eqref{eq:var} into Eq.~\eqref{eq:steady1}, with $\Lambda^\infty-1=\epsilon^2\,\chi_1^\infty$, and solving for $\chi_1^\infty$ gives
\begin{equation}
 \chi_1^\infty=\frac{2\,(\tau_1-1)}{1+\sqrt{1+4\,\alpha\,\epsilon^2\,(\tau_1-1)}}.
 \label{eq:closed}
\end{equation}
By Eq.~\eqref{eq:closed}, the weak-selection response holds when $\alpha\,\epsilon^2\,\tau_1$ is small, that is, when selection changes the growth rate by a small fraction. For strong selection, the response saturates, and the mean grows only as the square root of $\tau_1$ [Fig.~\ref{fig:steady}(b)].

To see how the modes set the best inheritance in a changing environment, suppose the environment switches periodically between two epochs, with selection fixed within each epoch. In the simplest case, both epochs last a time $T$, and the second reverses the selection profile of the first. For selection on a single mode, Eq.~\eqref{eq:Lambdamodes} then has a single term,
\begin{equation}
 s(x,t)=\sigma(t)\,s_i\,\psi_i(x),\quad
 \Lambda(t)=1+\epsilon^2\,\sigma(t)\,s_i\,\chi_i(t),
 \label{eq:sigma}
\end{equation}
where $\sigma(t)=+1$ in epoch $A$ and $\sigma(t)=-1$ in epoch $B$. Multiplying Eq.~\eqref{eq:modeexact} by $\epsilon^2\,\sigma\,s_i$, which is constant within an epoch, and using $\sigma^2=1$ gives the same equation in both epochs,
\begin{equation}
 \tau_i\,\dot\Lambda=\epsilon^2\,(\tau_i-1)\,s_i^2\Var(\psi_i)-\Lambda\,(\Lambda-1),
 \label{eq:Lambdadot}
\end{equation}
so the growth rate relaxes toward its steady value, given by Eq.~\eqref{eq:steadyexact}.

\begin{figure}[tb]
\includegraphics[width=\columnwidth]{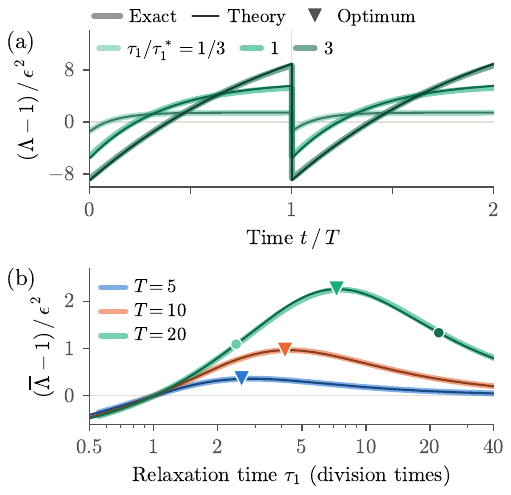}
\caption{Growth when selection reverses between $s(x)=x$ and $s(x)=-x$ every $T$ division times, for Gaussian inheritance and $\epsilon=0.1$. (a) Growth rate over one cycle for $T=20$ and relaxation times below, at, and above the optimum $\tau_1^*$, here the exact maximum of Eq.~\eqref{eq:opposite}. (b) Cycle-averaged growth rate against the relaxation time of the selected mode. Triangles mark the approximate optimum of Eq.~\eqref{eq:taustar}, and dots the relaxation times of (a). Thick lines: exact expected-density equation; thin lines: weak-selection theory, from Eq.~\eqref{eq:mode} in (a) and Eq.~\eqref{eq:opposite} in (b).}
\label{fig:periodic}
\end{figure}

At a switch, $\chi_i$ is continuous, because the population keeps its composition, but $\sigma$ and hence $\Lambda-1$ change sign: phenotypes enriched in one epoch divide slowly in the next [Fig.~\ref{fig:periodic}(a)]. Over a cycle, the changes of the growth rate within the epochs cancel its jumps at the switches, and averaging Eq.~\eqref{eq:Lambdadot}, with overlines for cycle averages, gives exactly (see SM)
\begin{equation}
\begin{aligned}
 \overline{\Lambda(\Lambda-1)}={}&\epsilon^2\,(\tau_i-1)\,s_i^2\,\overline{\Var(\psi_i)}\\
 &-\frac{\tau_i}{T}\,\big[\Lambda_A(T)+\Lambda_B(T)-2\big],
\end{aligned}
 \label{eq:cycle}
\end{equation}
where $\Lambda_A(T)$ and $\Lambda_B(T)$ are the growth rates at the ends of the two epochs. The subtracted term, the average of $\tau_i\,\dot\Lambda$ within the epochs, is the cost of switching.

For weak selection, the variance is one, and each mode reaches the fraction
\begin{equation}
 \theta_i\equiv\tanh(T/2\tau_i)
\end{equation}
of its steady response by the end of an epoch (see SM), so that $\Lambda_A(T)-1=\Lambda_B(T)-1=\epsilon^2\,(\tau_i-1)\,s_i^2\,\theta_i$. Inserting this into Eq.~\eqref{eq:cycle} gives
\begin{equation}
 \overline\Lambda\approx1+\epsilon^2\,(\tau_i-1)\,s_i^2\,\Big(1-\frac{2\,\tau_i\,\theta_i}{T}\Big).
 \label{eq:opposite}
\end{equation}

To find the best inheritance, we compare offspring distributions that have the same inheritance modes and differ only in their eigenvalues, so that $s_i$ is the same for all of them. Growth is then fastest at an intermediate relaxation time [Fig.~\ref{fig:periodic}(b)]: a mode much slower than an epoch spends each epoch undoing the response it built in the previous one, and a mode faster than one division responds against selection. For epochs much longer than a division time, the optimum is (see SM)
\begin{equation}
 \tau_i^*\approx\frac{2-(\theta^*)^2}{4\,\theta^*}\,T+\frac{2-(\theta^*)^2}{2\,\big[3\,(\theta^*)^2-2\big]},
 \label{eq:taustar}
\end{equation}
where $\theta^*$, the value of $\theta_i$ at the optimum to leading order, is the nonzero root of
\begin{equation}
 \theta^*=\tanh\frac{2\,\theta^*}{2-(\theta^*)^2}\approx0.9226.
 \label{eq:tstar}
\end{equation}
Numerically, $\tau_i^*\approx0.311\,T+1.04$: the selected mode should relax in about one third of an epoch, plus one division time. Because $\tau_i^*$ grows with $T$, these weak-selection results also require $\epsilon^2\,s_i^2\,T$ to be small (see SM). Inserting the optimum into Eq.~\eqref{eq:opposite} gives the optimal growth rate,
\begin{equation}
 \overline\Lambda\approx1+\epsilon^2\,s_i^2\bigg[\frac{2-(\theta^*)^2}{4\,\theta^*}\,T-1\bigg]\frac{(\theta^*)^2}{2},
 \label{eq:Lambdaopt}
\end{equation}
where the bracket is $\tau_i^*-1$ to leading order, so $(\theta^*)^2/2$ is the fraction of the growth-rate increase in a constant environment that survives in a periodic one. For weak selection, an intermediate optimum requires selection that is anticorrelated at some time lag, as when the environment reverses at regular intervals (see SM). If instead the autocorrelation of selection is nonnegative at all lags, as for epochs of exponentially distributed duration in a Markov chain, the gain of each mode increases monotonically with its relaxation time, and no intermediate relaxation time is best.

For Gaussian inheritance, only the eigenvalues, $\lambda_i=\rho^i$, depend on $\rho$, so inverting Eq.~\eqref{eq:tau} gives the optimal parent--offspring correlation, $(\rho^*)^i=1-1/(2\,\tau_i^*)$. Higher modes relax faster at the same $\rho$, so this optimal correlation increases with $i$. 
When selection acts on several modes, their weak-selection contributions add, each of the form of Eq.~\eqref{eq:opposite}, and the best positive $\rho$ is a compromise between their single-mode optima (see SM).

Molecular processes set the strength of inheritance, and regulation or mutations can tune it~\cite{Acar2005,Acar2008}. Stable proteins carry a cell's state into its daughters~\cite{Sigal2006,Lambert2014}, whereas random partitioning at division adds variation that weakens this memory~\cite{Huh2011}. Positive feedback~\cite{Acar2005,Kaufmann2007} and chromatin modifications~\cite{Bintu2016} can keep expression states over many generations. In the model, these processes set the offspring distribution and thus the relaxation times of the inheritance modes. Because these processes are genetically encoded, Eq.~\eqref{eq:taustar} predicts how inheritance should evolve in experiments with periodic environments~\cite{Beaumont2009}.

Selection in the model depends on the phenotype distribution only through the mean growth rate. In asexual genetic evolution, this is sufficient for the fitness distribution to travel as a wave~\cite{Tsimring1996,Rouzine2003,Desai2007,Hallatschek2011}. Here, non-genetic inheritance instead pulls the offspring distribution back toward $p^{(0)}$, bounding the response. If instead selection depended on the phenotype distribution directly, as under competition or predation~\cite{Jackson2024}, Eq.~\eqref{eq:modeexact} would still hold, but feedback could limit or amplify the response. Optimal inheritance would then have to be judged by whether a rare inheritance strategy can invade~\cite{Metz1992}. Division times in the model are exponentially distributed, whereas real ones vary much less~\cite{TaheriAraghi2015,Hashimoto2016}. Phenotypes can also change during a cell's lifetime~\cite{Pleska2021}, which would add relaxation independent of division. Cells that sense the environment could even regulate the size and the speed of a response independently, escaping the trade-off~\cite{Kussell2005,Wolf2005,Botero2015}. Without parent--offspring symmetry, $\mathcal P$ can be strongly non-normal, and observed responses can then combine large, partly cancelling contributions from modes far from orthogonal~\cite{Trefethen2005}. A phenotype can also have several variables, or traits, such as those that describe the swimming of \textit{E. coli}~\cite{Pleska2021,Vo2025}. Each trait can have its own parent--offspring correlation, and traits can be correlated with each other, both within a cell and between parent and offspring. The inheritance modes can then mix traits and need not align with the main axes of phenotypic variation.

Finally, non-genetic inheritance may change how much genetic diversity a community can support. In consumer--resource models of microbial communities, competitive exclusion allows no more species to coexist than there are resources~\cite{MacArthur1970,Levin1970,Advani2018,Marsland2019,Cui2020,Blumenthal2023,Li2026}. Many proposed mechanisms let more species coexist~\cite{Hutchinson1961,Armstrong1980,Huisman1999,Chesson2000,Scheffer2003,Kerr2002,Reichenbach2007,Posfai2017,Goldford2018,Goyal2018,Erez2020,Narla2025}, including environmental noise~\cite{AlHiyasat2026} and adaptive resource use~\cite{PaccianiMori2020}, whereas variation within species tends to hinder coexistence~\cite{Hart2016,Barabas2016}. Whether non-genetic inheritance helps or hinders coexistence is an open question.

\begin{acknowledgments}
\emph{Acknowledgements:} I thank Thierry Emonet for feedback on an early version of this work. I also thank Qiwei Yu for helpful discussions. The Flatiron Institute is a division of the Simons Foundation.

\emph{Data availability:} The data that support the findings of this article are available at \url{https://github.com/hhmattingly/inheritance_tradeoff}.
\end{acknowledgments}

\bibliography{references}

\end{document}


\title{Supplemental Material: Speed--response trade-off in non-genetic adaptation of a continuous phenotype}
\author{Henry Mattingly}
\affiliation{Center for Computational Biology, Flatiron Institute, New York, NY 10010, USA}
\maketitle
\renewcommand{\theequation}{S\arabic{equation}}
\renewcommand{\thefigure}{S\arabic{figure}}

\tableofcontents

\section{Model}
\label{sec:model}

We consider one genotype, so all phenotypic variation is non-genetic. Let the offspring distribution $\mathcal P(x'|x)$ be the probability density that an offspring of a parent with phenotype $x$ has phenotype $x'$. Every offspring has some phenotype, so
\begin{equation}
 \int \mathcal P(x'|x)\,dx'=1.
 \label{eq:normP}
\end{equation}
We assume that inheritance has a unique stationary distribution $p^{(0)}$, which parents pass on unchanged to their offspring,
\begin{equation}
 \int \mathcal P(x|y)\,p^{(0)}(y)\,dy=p^{(0)}(x),\qquad \int p^{(0)}(x)\,dx=1.
 \label{eq:p0}
\end{equation}
We measure the phenotype from its stationary mean in units of its stationary standard deviation,
\begin{equation}
 \int x\,p^{(0)}(x)\,dx=0,\qquad \int x^2\,p^{(0)}(x)\,dx=1.
 \label{eq:standard}
\end{equation}
This fixes the scale of the phenotype, and with it the meaning of $\epsilon$ for a selection profile given as a function of $x$, such as $s(x)=x$ below. Inheritance is parent--offspring symmetric (detailed balance, or reversibility) when the stationary joint distribution of parent and offspring phenotypes is unchanged by exchanging the two,
\begin{equation}
 p^{(0)}(y)\,\mathcal P(x|y)=p^{(0)}(x)\,\mathcal P(y|x).
 \label{eq:symmetric}
\end{equation}
We assume symmetric inheritance throughout. For discrete phenotypes, integrals become sums.

Let $n(x,t)$ be the expected population density. The population size and the phenotype distribution are
\begin{equation}
 N(t)\equiv\int n(x,t)\,dx,\qquad p(x,t)\equiv\frac{n(x,t)}{N(t)}.
 \label{eq:Np}
\end{equation}
In binary fission, a parent divides, disappears, and is replaced by two daughters, each of whose phenotypes is drawn from $\mathcal P$. The expected density depends only on the distribution of each daughter's phenotype, so correlations between sisters do not enter it. We measure time in units of the mean division time without selection, so that the neutral division rate is one, and let selection change the division rate of phenotype $x$ to $1+\epsilon\,s(x)$, where $\epsilon$ sets the strength of selection. We assume that this rate is nonnegative wherever the population lives. The expected density then obeys
\begin{equation}
 \partial_t n(x,t)=\int\big[2\,\mathcal P(x|y)-\delta(x-y)\big]\,\big(1+\epsilon\,s(y)\big)\,n(y,t)\,dy.
 \label{eq:fission}
\end{equation}
Each division of a parent with phenotype $y$, which occurs at rate $1+\epsilon\,s(y)$, adds two daughters whose phenotypes are drawn from $\mathcal P$ and removes the parent itself, which the delta function accounts for.

Each division adds one individual net. Integrating Eq.~\eqref{eq:fission} over $x$ therefore shows that the population growth rate equals the mean division rate,
\begin{equation}
 \Lambda(t)\equiv\frac{\dot N(t)}{N(t)}=1+\epsilon\int s(x)\,p(x,t)\,dx.
 \label{eq:Lambda}
\end{equation}
Differentiating $p=n/N$ gives the exact equation for the phenotype distribution,
\begin{equation}
 \partial_t p(x,t)=\int\big[2\,\mathcal P(x|y)-\delta(x-y)\big]\,\big(1+\epsilon\,s(y)\big)\,p(y,t)\,dy-\Lambda(t)\,p(x,t),
 \label{eq:p}
\end{equation}
where the last term keeps the distribution normalized.
A part of $s$ that is the same for all phenotypes only changes the overall division rate. Since $1+\epsilon\,[s(x)+c]=(1+\epsilon\,c)\,[1+\epsilon\,s(x)/(1+\epsilon\,c)]$ for a constant $c$, it can be absorbed exactly into the unit of time and the strength of selection. We therefore center the selection profile,
\begin{equation}
 \int s(x)\,p^{(0)}(x)\,dx=0.
 \label{eq:center}
\end{equation}

\section{Inheritance modes}
\label{sec:modes}

We describe the phenotype distribution through inheritance modes. The modes $\psi_i$ are functions of the phenotype whose expected value in an offspring is proportional to their value in the parent,
\begin{equation}
 \int \mathcal P(x'|x)\,\psi_i(x')\,dx'=\lambda_i\,\psi_i(x).
 \label{eq:left}
\end{equation}
They are the left eigenfunctions of $\mathcal P$. The constant function is a mode with eigenvalue one. We label it by $i=0$. Because the stationary joint distribution of parent and offspring is symmetric, the eigenvalue problem of Eq.~\eqref{eq:left} is self-adjoint with weight $p^{(0)}$. The eigenvalues are therefore real, and the modes can be chosen orthonormal with weight $p^{(0)}$,
\begin{equation}
 \int p^{(0)}(x)\,\psi_i(x)\,\psi_j(x)\,dx=\delta_{ij}.
 \label{eq:ortho}
\end{equation}
Orthogonality to the constant function means that every mode other than the stationary one averages to zero over $p^{(0)}$. Because $p^{(0)}$ is the unique stationary distribution, all other eigenvalues are smaller than one. We assume that the modes form a complete set. This holds for finite phenotype spaces and for offspring distributions that define compact integral operators, such as the Gaussian one of Sec.~\ref{sec:gaussian}. Repeated mode indices are not summed. All sums are written explicitly.

Selection moves the phenotype distribution away from $p^{(0)}$. By completeness, the ratio $p/p^{(0)}$ can be expanded in the modes. Its coefficient for the constant mode is one, because $p$ is normalized, and we write the expansion as
\begin{equation}
 p(x,t)=p^{(0)}(x)\,\Big[1+\epsilon\sum_{i\geq1}\chi_i(t)\,\psi_i(x)\Big],
 \label{eq:expand}
\end{equation}
which defines the mode amplitudes $\chi_i$. The sum is the relative change of the distribution caused by selection, and $\chi_i$ measures how much of mode $i$ it contains. The factor $\epsilon$ keeps the amplitudes finite as selection weakens. Without selection, every amplitude vanishes. To express the amplitudes through the distribution, we denote averages over the current population by
\begin{equation}
 \langle f\rangle\equiv\int f(x)\,p(x,t)\,dx
 \label{eq:avg}
\end{equation}
for any function $f$ of the phenotype. Multiplying Eq.~\eqref{eq:expand} by $\psi_i(x)$, integrating over $x$, and using Eq.~\eqref{eq:ortho} shows that each amplitude is the average of its mode,
\begin{equation}
 \chi_i(t)=\frac{1}{\epsilon}\,\langle\psi_i\rangle.
 \label{eq:chii}
\end{equation}
Likewise, the selection coefficients are the coefficients of the selection profile in the modes,
\begin{equation}
 s_i\equiv\int p^{(0)}(x)\,\psi_i(x)\,s(x)\,dx,\qquad
 s(x)=\sum_{i\geq1}s_i\,\psi_i(x),
 \label{eq:si}
\end{equation}
where the constant mode is absent because $s$ is centered. The selection coefficient $s_i$ measures how strongly selection acts on mode $i$. Inserting Eq.~\eqref{eq:expand} into Eq.~\eqref{eq:Lambda} and using Eqs.~\eqref{eq:center}, \eqref{eq:ortho}, and \eqref{eq:si} gives the growth rate exactly as a sum over modes,
\begin{equation}
 \Lambda(t)=1+\epsilon^2\sum_{i\geq1}s_i\,\chi_i(t).
 \label{eq:Lambdamodes}
\end{equation}
Each mode affects growth through its own selection coefficient.

\subsection{Exact mode equation}
\label{sec:modeexact}
To obtain the dynamics of the amplitudes, we multiply Eq.~\eqref{eq:p} by $\psi_i(x)$ and integrate over $x$. In the term with $\mathcal P$, Eq.~\eqref{eq:left} replaces $\int\psi_i(x)\,\mathcal P(x|y)\,dx$ by $\lambda_i\,\psi_i(y)$, and the delta function gives $\psi_i(y)$, so
\begin{equation}
 \frac{d}{dt}\langle\psi_i\rangle=(2\lambda_i-1)\,\big\langle\psi_i\,(1+\epsilon\,s)\big\rangle-\Lambda(t)\,\langle\psi_i\rangle.
 \label{eq:modestep}
\end{equation}
The coefficient $2\lambda_i-1$ counts the two daughters, whose content of mode $i$ is multiplied by $\lambda_i$ by inheritance, minus the parent they replace. By Eq.~\eqref{eq:Lambda}, the mean division rate is $\langle1+\epsilon\,s\rangle=\Lambda$, so the first average on the right splits into a part that follows the mode and a covariance,
\begin{equation}
 \big\langle\psi_i\,(1+\epsilon\,s)\big\rangle=\Lambda(t)\,\langle\psi_i\rangle+\epsilon\Cov(\psi_i,s),
\end{equation}
where the covariance and the variance over the current population are
\begin{equation}
 \Cov(f,g)\equiv\langle f\,g\rangle-\langle f\rangle\langle g\rangle,\qquad
 \Var(f)\equiv\Cov(f,f).
 \label{eq:covdef}
\end{equation}
Inserting this into Eq.~\eqref{eq:modestep} and dividing by $\epsilon$ gives
\begin{equation}
 \dot \chi_i(t)=-2\,(1-\lambda_i)\,\Lambda(t)\,\chi_i(t)+(2\lambda_i-1)\Cov(\psi_i,s).
 \label{eq:mode_lambda}
\end{equation}
The relaxation time of mode $i$ is
\begin{equation}
 \tau_i\equiv\frac{1}{2\,(1-\lambda_i)},
 \label{eq:tau}
\end{equation}
where the factor two counts the two inheritance events per division. Eliminating $\lambda_i$ in favor of $\tau_i$, the mode equation becomes
\begin{equation}
 \tau_i\,\dot \chi_i(t)=(\tau_i-1)\Cov(\psi_i,s)-\Lambda(t)\,\chi_i(t).
 \label{eq:modeexact}
\end{equation}

Equation~\eqref{eq:modeexact} is exact for any selection strength. It also holds for asymmetric inheritance, for which the modes are not orthogonal. Equation~\eqref{eq:chii} then no longer follows from Eq.~\eqref{eq:expand}, but Eq.~\eqref{eq:modeexact} still holds for $\chi_i$ defined as $\langle\psi_i\rangle/\epsilon$, because its derivation used only the left eigenvalue relation, Eq.~\eqref{eq:left}. Each mode relaxes on the timescale $\tau_i$ and is forced by its covariance with the selection profile in the current population, as in Price's equation~\cite{Price1970}. Without selection, $\Lambda=1$, and Eq.~\eqref{eq:modeexact}, multiplied by $\epsilon$, shows that the average of each mode decays exponentially on the timescale $\tau_i$. Selection enters in two further ways, both of which couple the modes. First, it changes the mean division rate to $\Lambda$, which depends on all amplitudes through Eq.~\eqref{eq:Lambdamodes}, and with it how often phenotypes are redrawn: every mode relaxes faster by the factor $\Lambda$ when $\Lambda>1$ and more slowly when $\Lambda<1$. Second, the covariance changes as the population responds. Using Eqs.~\eqref{eq:expand} and \eqref{eq:si},
\begin{equation}
 \Cov(\psi_i,s)=s_i+\epsilon\sum_{k\geq1}\chi_k(t)\int p^{(0)}(x)\,\psi_i(x)\,\psi_k(x)\,s(x)\,dx-\epsilon^2\,\chi_i(t)\sum_{j\geq1}s_j\,\chi_j(t),
 \label{eq:cov}
\end{equation}
where the last term is the product of $\langle\psi_i\rangle$ and $\langle s\rangle$, the latter given by Eqs.~\eqref{eq:Lambda} and \eqref{eq:Lambdamodes}. The middle term couples mode $i$ to the other modes through the selection profile, and the last term through the growth rate.

\subsection{Gaussian inheritance}
\label{sec:gaussian}

As an example, let the offspring phenotype regress toward the stationary mean with parent--offspring correlation $\rho$, which lies strictly between $-1$ and $1$. In the units of Eq.~\eqref{eq:standard}, an offspring of a parent with phenotype $x$ has phenotype
\begin{equation}
 x'=\rho\,x+z,
 \label{eq:gaussian}
\end{equation}
where the noise $z$ is independent of $x$ and normally distributed, with mean zero and variance $1-\rho^2$. The offspring distribution and the stationary distribution are
\begin{equation}
 \mathcal P(x'|x)=\mathcal N\big(\rho\,x,\;1-\rho^2\big),\qquad
 p^{(0)}(x)=\mathcal N(0,\;1),
 \label{eq:gausskernel}
\end{equation}
where $\mathcal N$ denotes a normal density with the given mean and variance, evaluated at $x'$ and at $x$, respectively.
At stationarity, parent and offspring phenotypes are jointly normal with unit variances and correlation $\rho$. This joint law is unchanged by exchanging them, so inheritance is parent--offspring symmetric.

The modes are the normalized Hermite polynomials, the probabilists' Hermite polynomials divided by the square root of $i!$,
\begin{equation}
 \psi_i(x)=\frac{(-1)^i}{\sqrt{i!}}\,e^{x^2/2}\,\frac{d^i}{dx^i}\,e^{-x^2/2},\qquad
 \psi_0(x)=1,\quad \psi_1(x)=x,\quad \psi_2(x)=\frac{x^2-1}{\sqrt2}.
 \label{eq:hermitedef}
\end{equation}
They are orthonormal with weight $p^{(0)}$, as Eq.~\eqref{eq:ortho} requires, and the expected value of $\psi_i$ in an offspring is $\rho^i$ times its value in the parent, as follows from the generating function of the Hermite polynomials (Mehler's formula~\cite{Mehler1866}). The eigenvalues and, by Eq.~\eqref{eq:tau}, the relaxation times are therefore
\begin{equation}
 \lambda_i=\rho^i,\qquad
 \tau_i=\frac{1}{2\,(1-\rho^i)}.
 \label{eq:hermite}
\end{equation}
The corresponding changes of the distribution, $p^{(0)}$ times $\psi_i$, are proportional to derivatives of $p^{(0)}$. The first mode shifts the mean of the phenotype distribution, the second changes its variance, and higher modes change higher moments. The selection coefficients $s_i$ are the coefficients of the selection profile in Hermite polynomials, Eq.~\eqref{eq:si}. A linear profile, which describes directional selection, acts only on the first mode, and a quadratic profile symmetric about the stationary mean, which describes stabilizing or disruptive selection, only on the second. Multiplication by the phenotype raises or lowers the index of a Hermite polynomial by one,
\begin{equation}
 x\,\psi_i(x)=\sqrt{i+1}\,\psi_{i+1}(x)+\sqrt{i}\,\psi_{i-1}(x),
 \label{eq:recurrence}
\end{equation}
so that, for example, $x^2=1+\sqrt2\,\psi_2(x)$. By Eq.~\eqref{eq:chii}, the mode amplitudes are averages of Hermite polynomials and therefore combinations of the moments of the phenotype distribution: the amplitude of mode $i$ involves the moments up to the $i$th. The first two amplitudes give the mean and the variance,
\begin{equation}
 \langle x\rangle=\epsilon\,\chi_1,\qquad
 \Var(x)=\langle x^2\rangle-\langle x\rangle^2=1+\sqrt2\,\epsilon\,\chi_2-\epsilon^2\,\chi_1^2.
 \label{eq:moments}
\end{equation}

For positive $\rho$, the first mode, which shifts the mean, is the slowest. For negative $\rho$, the largest nonconstant eigenvalue is $\rho^2$, so the second mode is the slowest. Because $\lambda_1=\rho<0$, the mean then relaxes faster than for uncorrelated inheritance. For uncorrelated inheritance, every mode relaxes in half a division time.

\section{Speed--response trade-off}
\label{sec:tradeoff}

In a constant environment, the population approaches a steady state, in which the phenotype distribution no longer changes and the population grows at the rate $\Lambda^\infty$. A superscript $\infty$ marks values in the steady population. Setting the time derivative in Eq.~\eqref{eq:modeexact} to zero gives
\begin{equation}
 \Lambda^\infty\,\chi_i^\infty=(\tau_i-1)\Cov^\infty(\psi_i,s).
 \label{eq:steadyexact}
\end{equation}
Multiplying by $\epsilon^2\,s_i$, summing over $i$, and using Eq.~\eqref{eq:Lambdamodes} on the left and the expansion of $s$ in Eq.~\eqref{eq:si} inside the covariance gives the steady growth rate,
\begin{equation}
 \Lambda^\infty\,\big(\Lambda^\infty-1\big)=\epsilon^2\sum_{i\geq1}(\tau_i-1)\,s_i\Cov^\infty(\psi_i,s)
 =\epsilon^2\sum_{i,j\geq1}(\tau_i-1)\,s_i\,s_j\Cov^\infty(\psi_i,\psi_j).
 \label{eq:Lambdasteady}
\end{equation}
Both relations are exact for any selection profile. The first holds for any offspring distribution, and the second, which uses the orthonormality of the modes through Eq.~\eqref{eq:Lambdamodes}, only for symmetric inheritance. In the stationary population the modes are uncorrelated and have unit variance, by Eq.~\eqref{eq:ortho}. Selection changes their variances and correlates them, so that each mode also responds to selection on the others. This is the coupling of Eq.~\eqref{eq:cov}.

\subsection{Weak selection}
\label{sec:weak}
For weak selection, the covariances take their stationary values: by Eq.~\eqref{eq:cov}, the covariance in Eq.~\eqref{eq:modeexact} reduces to $s_i$, with corrections of order $\epsilon$, and by Eq.~\eqref{eq:Lambdamodes} the growth rate reduces to one, with corrections of order $\epsilon^2$. To leading order, Eq.~\eqref{eq:modeexact} becomes
\begin{equation}
 \tau_i\,\dot \chi_i(t)=(\tau_i-1)\,s_i-\chi_i(t).
 \label{eq:mode}
\end{equation}
Each mode amplitude is then driven only by its own selection coefficient and relaxes on the timescale $\tau_i$ toward the steady response
\begin{equation}
 \chi_i^\infty=(\tau_i-1)\,s_i.
 \label{eq:chiinf}
\end{equation}
This is the speed--response trade-off: at a fixed selection coefficient, the more slowly a mode relaxes, the larger its steady response. The response has the sign of $s_i$ only for modes that relax more slowly than one division time, that is, for eigenvalues above one half. Faster modes respond against selection: favored cells divide more often, and each division replaces a parent by two daughters whose values of $\psi_i$ add up, on average, to only $2\lambda_i$ times its own.

In a constant environment, Eqs.~\eqref{eq:Lambdamodes} and \eqref{eq:chiinf} give the steady growth rate
\begin{equation}
 \Lambda^\infty=1+\epsilon^2\sum_{i\geq1}(\tau_i-1)\,s_i^2+O(\epsilon^3),
 \label{eq:Lambdastar}
\end{equation}
to which modes with eigenvalues below one half contribute negatively. It is also the weak-selection limit of Eq.~\eqref{eq:Lambdasteady}. Section~\ref{sec:mean} shows that, for slowly relaxing modes, the weak-selection limit requires $\epsilon^2\,\tau_i$ to be small, not only $\epsilon$.

\subsection{Beyond weak selection: directional selection}
\label{sec:mean}
When selection acts on a single mode, $s(x)=s_i\,\psi_i(x)$, the covariance in Eq.~\eqref{eq:steadyexact} is $s_i$ times the variance of that mode, so the steady response is
\begin{equation}
 \chi_i^\infty=\frac{(\tau_i-1)\,s_i\Var^\infty(\psi_i)}{\Lambda^\infty}.
 \label{eq:steadysingle}
\end{equation}
By Eq.~\eqref{eq:ortho}, this variance is one in the stationary population, and for weak selection $\Lambda^\infty$ is close to one, which recovers Eq.~\eqref{eq:chiinf}. Beyond weak selection, the response therefore changes in two ways. Selection changes the growth rate, and with it how often phenotypes are redrawn, which divides the response by $\Lambda^\infty$: relaxation is faster when $\Lambda^\infty>1$ and slower when $\Lambda^\infty<1$. This holds for every offspring distribution. And selection changes the variance of the selected mode, which multiplies the response. This variance depends on the offspring distribution, and we evaluate it for Gaussian inheritance.

Let $s(x)=x$. This is $\psi_1$, so selection acts on the first mode alone, with $s_1=1$, and by Eqs.~\eqref{eq:Lambdamodes} and \eqref{eq:moments} the growth rate is
\begin{equation}
 \Lambda(t)=1+\epsilon^2\,\chi_1(t)=1+\epsilon\,\langle x\rangle.
 \label{eq:Lambdamean}
\end{equation}
The covariances in Eq.~\eqref{eq:modeexact} for the first two modes are covariances with $x$ of $x$ and of $\psi_2=(x^2-1)/\sqrt2$. The first is the variance. The second involves the third moment,
\begin{equation}
 \Cov(\psi_2,x)=\frac{\Cov(x^2,x)}{\sqrt2}=\sqrt2\,\langle x\rangle\Var(x)+\frac{\kappa_3}{\sqrt2},\qquad
 \kappa_3\equiv\Big\langle\big(x-\langle x\rangle\big)^3\Big\rangle,
 \label{eq:kappa3}
\end{equation}
where $\kappa_3$ is the third cumulant, which vanishes for a normal distribution. With $\langle x\rangle=\epsilon\,\chi_1$, Eq.~\eqref{eq:modeexact} for the first two modes is therefore
\begin{equation}
\begin{aligned}
 \tau_1\,\dot \chi_1(t)&=(\tau_1-1)\Var(x)-\Lambda(t)\,\chi_1(t),\\
 \tau_2\,\dot \chi_2(t)&=(\tau_2-1)\Big[\sqrt2\,\epsilon\,\chi_1(t)\Var(x)+\frac{\kappa_3}{\sqrt2}\Big]-\Lambda(t)\,\chi_2(t),
\end{aligned}
 \label{eq:chi12exact}
\end{equation}
which is exact. These equations do not close: through $\kappa_3$, the second involves the third mode. Selection multiplies the distribution by $x$, which by Eq.~\eqref{eq:recurrence} raises the mode index, so the equation for each mode involves the next. The higher modes are, however, of higher order in $\epsilon$. In Eq.~\eqref{eq:modeexact} for mode $i$, multiplied by $\epsilon$, the forcing is $\epsilon$ times the covariance of $\psi_i$ with $x$, which by Eq.~\eqref{eq:recurrence} involves the averages of $\psi_{i+1}$ and $\psi_{i-1}$. Expanding in $\epsilon$ and starting from $\langle\psi_0\rangle=1$, the average of $\psi_i$, which is $\epsilon\,\chi_i$, therefore first appears at order $\epsilon^i$. The third cumulant, which involves $\langle\psi_3\rangle$ and products of lower moments, is then of order $\epsilon^3$. By Eq.~\eqref{eq:chi12exact}, it changes $\chi_2$ at order $\epsilon^3$, hence the variance at order $\epsilon^4$ by Eq.~\eqref{eq:moments}, $\chi_1$ at order $\epsilon^4$, and the growth rate at order $\epsilon^6$ by Eq.~\eqref{eq:Lambdamean}.

We neglect the third cumulant, which closes Eq.~\eqref{eq:chi12exact},
\begin{equation}
\begin{aligned}
 \tau_1\,\dot \chi_1(t)&=(\tau_1-1)\Var(x)-\Lambda(t)\,\chi_1(t),\\
 \tau_2\,\dot \chi_2(t)&=\sqrt2\,\epsilon\,(\tau_2-1)\,\chi_1(t)\Var(x)-\Lambda(t)\,\chi_2(t),
\end{aligned}
 \label{eq:chi12}
\end{equation}
with the variance from Eq.~\eqref{eq:moments} and the growth rate from Eq.~\eqref{eq:Lambdamean}. Neglecting the third cumulant is equivalent to approximating the phenotype distribution by a normal distribution with the current mean and variance. Integrating Eq.~\eqref{eq:chi12} from the stationary state, $\chi_1=\chi_2=0$, gives the transient response after selection is switched on.

In the steady state, Eq.~\eqref{eq:chi12} gives
\begin{equation}
 \Lambda^\infty\,\chi_1^\infty=(\tau_1-1)\Var^\infty(x),\qquad
 \Lambda^\infty\,\chi_2^\infty=\sqrt2\,\epsilon\,(\tau_2-1)\,\chi_1^\infty\Var^\infty(x).
 \label{eq:steadyamps}
\end{equation}
The first relation is Eq.~\eqref{eq:steadysingle} for this profile, and it is exact, because the third cumulant enters only the second. To find the variance, we insert both steady amplitudes into Eq.~\eqref{eq:moments}. Using $\epsilon^2\,\chi_1^\infty=\Lambda^\infty-1$, the two terms that depend on selection are
\begin{equation}
 \sqrt2\,\epsilon\,\chi_2^\infty=2\,(\tau_2-1)\,\frac{\Lambda^\infty-1}{\Lambda^\infty}\Var^\infty(x),\qquad
 \epsilon^2\,\big(\chi_1^\infty\big)^2=(\tau_1-1)\,\frac{\Lambda^\infty-1}{\Lambda^\infty}\Var^\infty(x),
\end{equation}
so that Eq.~\eqref{eq:moments} becomes
\begin{equation}
 \Var^\infty(x)=1+(2\,\tau_2-\tau_1-1)\,\frac{\Lambda^\infty-1}{\Lambda^\infty}\Var^\infty(x).
\end{equation}
Its solution is the steady variance,
\begin{equation}
 \Var^\infty(x)=\frac{\Lambda^\infty}{1+\alpha\,\big(\Lambda^\infty-1\big)},
 \label{eq:var}
\end{equation}
with
\begin{equation}
 \alpha\equiv\tau_1-2\,\tau_2+2=2-\frac{1}{2\,(1+\rho)},
 \label{eq:alpha}
\end{equation}
where the second form follows from Eq.~\eqref{eq:hermite}. The variance differs from one by $(1-\alpha)\,(\Lambda^\infty-1)/[1+\alpha\,(\Lambda^\infty-1)]$. For $\rho>-1/2$, $\alpha$ exceeds one, so the population narrows when selection raises the growth rate, as it does for modes that relax more slowly than one division time, and widens when selection lowers it.

Inserting Eq.~\eqref{eq:var} into the first relation of Eq.~\eqref{eq:steadyamps}, with $\Lambda^\infty-1=\epsilon^2\,\chi_1^\infty$ from Eq.~\eqref{eq:Lambdamean}, gives a quadratic equation for the steady response of the first mode,
\begin{equation}
 \chi_1^\infty=\frac{\tau_1-1}{1+\alpha\,\epsilon^2\,\chi_1^\infty}.
 \label{eq:chi1closed}
\end{equation}
Its root that reduces to the weak-selection response as $\epsilon$ goes to zero is
\begin{equation}
 \chi_1^\infty=\frac{2\,(\tau_1-1)}{1+\sqrt{1+4\,\alpha\,\epsilon^2\,(\tau_1-1)}}.
 \label{eq:closed}
\end{equation}

Strong selection thus divides the weak-selection response, Eq.~\eqref{eq:chiinf}, by the factor $1+\alpha\,\epsilon^2\,\chi_1^\infty$, which by Eq.~\eqref{eq:var} equals $\Lambda^\infty/\Var^\infty(x)$. For modes that relax more slowly than one division time, $\Lambda^\infty>1$, and the response is reduced in the two ways of Eq.~\eqref{eq:steadysingle}: faster relaxation divides it by $\Lambda^\infty$, and narrowing multiplies it by $\Var^\infty(x)<1$. To first order in $\Lambda^\infty-1$, the first contributes one unit of $\alpha$ and the second $\alpha-1$, since $\Var^\infty(x)\approx1-(\alpha-1)\,(\Lambda^\infty-1)$. For positive $\rho$, $\alpha$ lies between $3/2$ and $7/4$, so the two effects are comparable. For faster modes, $\Lambda^\infty<1$, so relaxation slows, and for $\rho>-1/2$ the distribution also widens. Both enlarge the negative response. Expanding Eq.~\eqref{eq:closed} in $\epsilon$,
\begin{equation}
 \Lambda^\infty=1+\epsilon^2\,(\tau_1-1)-\alpha\,\epsilon^4\,(\tau_1-1)^2+O(\epsilon^6).
 \label{eq:Lambdaexpand}
\end{equation}
The coefficient of $\epsilon^4$ is exact, because the neglected third cumulant first affects the growth rate at order $\epsilon^6$. Odd orders vanish, because reversing the sign of the phenotype leaves the Gaussian offspring distribution unchanged and reverses the selection profile, which is equivalent to reversing the sign of $\epsilon$.

The weak-selection response is therefore accurate when $\alpha\,\epsilon^2\,\tau_1$ is small. For slowly relaxing modes, this condition is stronger than a small $\epsilon$, but weaker than requiring the shift of the mean, about $\epsilon\,\tau_1$ standard deviations, to be small: the relevant small parameter is the relative change of the growth rate. For strong selection, where $4\,\alpha\,\epsilon^2\,(\tau_1-1)$ is large, Eq.~\eqref{eq:closed} gives a steady mean phenotype of about $\sqrt{(\tau_1-1)/\alpha}$, which grows only as the square root of the relaxation time.

Written for the mean and the second moment, Eq.~\eqref{eq:chi12exact} uses only that $x$ and $x^2-1$ are left eigenfunctions of $\mathcal P$ with eigenvalues $\rho$ and $\rho^2$. This holds whenever the offspring phenotype is $\rho$ times the parent's plus independent noise: if the noise $z$ in Eq.~\eqref{eq:gaussian} has mean zero and variance $1-\rho^2$ but is otherwise arbitrarily distributed, then
\begin{equation}
 \int\mathcal P(x'|x)\,x'\,dx'=\rho\,x,\qquad
 \int\mathcal P(x'|x)\,\big(x'^2-1\big)\,dx'=\rho^2\,\big(x^2-1\big).
\end{equation}
Such inheritance is in general not parent--offspring symmetric, which the rest of this Supplemental Material assumes. The results of this subsection hold nonetheless, because none of the relations they use requires symmetry. The mode equation, Eq.~\eqref{eq:modeexact}, holds without symmetry, as shown in Sec.~\ref{sec:modeexact}. Equation~\eqref{eq:moments} uses only the definitions of $\psi_1$ and $\psi_2$. And Eq.~\eqref{eq:Lambdamean} follows directly from Eq.~\eqref{eq:Lambda} for $s(x)=x$. The normal distribution enters only through the neglect of the third cumulant, which for other noise distributions is generally not of third order in $\epsilon$.

\section{Growth in a periodically changing environment}
\label{sec:switching}

Let the environment switch periodically between two epochs, $A$ and $B$, of durations $T_A$ and $T_B$, with centered selection profiles $s^A$ and $s^B$ and selection coefficients $s_i^A$ and $s_i^B$. We consider the periodic state that the population reaches after many cycles, and the growth rate averaged over a cycle,
\begin{equation}
 \overline\Lambda\equiv\frac{1}{T_A+T_B}\int_0^{T_A+T_B}\Lambda(t)\,dt.
 \label{eq:cycleavg}
\end{equation}
Except in Sec.~\ref{sec:switchstrong}, which is exact, this section and Sec.~\ref{sec:optimum} use the weak-selection limit, Eq.~\eqref{eq:mode}.

Consider one mode and drop its index. Let $\chi_A(t)$ and $\chi_B(t)$ be its amplitude during epochs $A$ and $B$, with $t$ measured from the start of each epoch, and let $\chi_A^\infty$ and $\chi_B^\infty$ be the steady responses of Eq.~\eqref{eq:chiinf} in the two epochs. Within epoch $A$, Eq.~\eqref{eq:mode} gives
\begin{equation}
 \chi_A(t)=\chi_A^\infty+\big[\chi_A(0)-\chi_A^\infty\big]\,e^{-t/\tau},
 \label{eq:epoch}
\end{equation}
and likewise in epoch $B$. Each epoch starts where the previous one ended,
\begin{equation}
 \chi_B(0)=\chi_A(T_A),\qquad \chi_A(0)=\chi_B(T_B).
 \label{eq:matching}
\end{equation}
Solving these two conditions gives the lag behind the steady response at the start of epoch $A$,
\begin{equation}
 \chi_A(0)-\chi_A^\infty=-\big(\chi_A^\infty-\chi_B^\infty\big)\,\frac{1-e^{-T_B/\tau}}{1-e^{-(T_A+T_B)/\tau}},
 \label{eq:lagstart}
\end{equation}
and the same expression with $A$ and $B$ exchanged at the start of epoch $B$. Integrating Eq.~\eqref{eq:epoch} over the epoch,
\begin{equation}
 \int_0^{T_A}\chi_A(t)\,dt=\chi_A^\infty\,T_A-\big(\chi_A^\infty-\chi_B^\infty\big)\,L=\chi_B^\infty\,L+\chi_A^\infty\,(T_A-L),
 \label{eq:epochint}
\end{equation}
where
\begin{equation}
 L\equiv\tau\,\frac{\big(1-e^{-T_A/\tau}\big)\big(1-e^{-T_B/\tau}\big)}{1-e^{-(T_A+T_B)/\tau}}
 \label{eq:L}
\end{equation}
is symmetric in the two epochs. We call $L$ the switching lag: by the second form of Eq.~\eqref{eq:epochint}, the epoch contributes as if the amplitude stayed at the previous steady response for a time $L$ after the switch and then jumped to the new one. For long epochs, $L$ approaches $\tau$.

Restoring the mode index, writing $L_i$ for Eq.~\eqref{eq:L} with $\tau$ replaced by $\tau_i$, and using Eqs.~\eqref{eq:Lambdamodes} and \eqref{eq:chiinf} in Eq.~\eqref{eq:cycleavg} gives
\begin{equation}
 \overline\Lambda=1+\frac{\epsilon^2}{T_A+T_B}\sum_{i\geq1}(\tau_i-1)\Big[T_A\,(s_i^A)^2+T_B\,(s_i^B)^2-L_i\,(s_i^A-s_i^B)^2\Big]+O(\epsilon^3).
 \label{eq:periodic}
\end{equation}
The first two terms in the bracket are the steady gains of Eq.~\eqref{eq:Lambdastar} in the two epochs, weighted by their durations. The last is the cost of lagging behind after the two switches. No assumption on the epoch durations has been made, and each mode contributes independently.

\subsection{Opposite profiles in equal epochs}
\label{sec:opposite}
The simplest case is an environment in which epoch $B$ reverses the selection profile of epoch $A$ and both epochs last the same time $T$,
\begin{equation}
 s^B(x)=-s^A(x),\qquad T_A=T_B\equiv T.
 \label{eq:oppositecase}
\end{equation}
The selection coefficients of the two epochs are then opposite, and we write $s_i$ for that of epoch $A$,
\begin{equation}
 s_i\equiv s_i^A=-s_i^B.
\end{equation}
In the periodic state, Eq.~\eqref{eq:lagstart} with $\chi_B^\infty=-\chi_A^\infty$ gives $\chi_A(0)=-\theta_i\,\chi_A^\infty$, and, since $\chi_B(t)=-\chi_A(t)$ by symmetry, Eq.~\eqref{eq:matching} gives $\chi_A(T)=-\chi_A(0)=\theta_i\,\chi_A^\infty$, where
\begin{equation}
 \theta_i\equiv\tanh\frac{T}{2\tau_i}.
 \label{eq:thetadef}
\end{equation}
Each mode thus reaches the fraction $\theta_i$ of its steady response, Eq.~\eqref{eq:chiinf}, by the end of an epoch. The switching lag of Eq.~\eqref{eq:L} becomes $L_i=\tau_i\,\theta_i$, and Eq.~\eqref{eq:periodic} reduces to
\begin{equation}
 \overline\Lambda=1+\epsilon^2\sum_{i\geq1}(\tau_i-1)\,s_i^2\,F(\tau_i)+O(\epsilon^3),
 \label{eq:opposite}
\end{equation}
where
\begin{equation}
 F(\tau_i)\equiv1-\frac{2\,\tau_i\,\theta_i}{T}.
 \label{eq:F}
\end{equation}
The factor $F$ lies between zero and one. It approaches one for modes that relax much faster than an epoch lasts. For modes that relax much more slowly, it vanishes as one twelfth of the square of the ratio of epoch duration to relaxation time, because these modes spend each epoch undoing the response built in the previous one. Each mode thus contributes its steady gain, reduced by the finite epoch duration. For a single mode, this is the opposite-profile result of the main text.

\subsection{Exact growth-rate identity}
\label{sec:switchstrong}
We now return to the exact mode equation, Eq.~\eqref{eq:modeexact}. Let selection act on a single mode $i$, with opposite profiles in the two epochs and arbitrary epoch durations,
\begin{equation}
 s(x,t)=\sigma(t)\,s_i\,\psi_i(x),\qquad
 \sigma(t)=\begin{cases}+1&\text{in epoch $A$},\\-1&\text{in epoch $B$}.\end{cases}
 \label{eq:sigma}
\end{equation}
By Eqs.~\eqref{eq:Lambda} and \eqref{eq:chii}, the growth rate is
\begin{equation}
 \Lambda(t)=1+\epsilon^2\,\sigma(t)\,s_i\,\chi_i(t),
 \label{eq:Lambdasigma}
\end{equation}
and the covariance in Eq.~\eqref{eq:modeexact} is $\sigma(t)\,s_i\Var(\psi_i)$. Within an epoch, $\sigma$ is constant, so multiplying Eq.~\eqref{eq:modeexact} by $\epsilon^2\,\sigma(t)\,s_i$ turns its left side into $\tau_i\,\dot\Lambda$ and gives an exact equation for the growth rate,
\begin{equation}
 \tau_i\,\dot\Lambda(t)=\epsilon^2\,(\tau_i-1)\,s_i^2\Var(\psi_i)-\Lambda(t)\,\big[\Lambda(t)-1\big].
 \label{eq:Lambdadot}
\end{equation}
In a constant environment, its steady state is equivalent to Eq.~\eqref{eq:steadysingle}. At a switch, the amplitude $\chi_i$ is continuous but $\sigma$ changes sign, so by Eq.~\eqref{eq:Lambdasigma} the excess growth rate $\Lambda-1$ changes sign. Denoting the growth rates during the two epochs by $\Lambda_A(t)$ and $\Lambda_B(t)$, with $t$ measured from the start of each, as in Eq.~\eqref{eq:epoch},
\begin{equation}
 \Lambda_B(0)-1=-\big[\Lambda_A(T_A)-1\big],\qquad
 \Lambda_A(0)-1=-\big[\Lambda_B(T_B)-1\big].
 \label{eq:Lambdajump}
\end{equation}
We integrate Eq.~\eqref{eq:Lambdadot} over the two epochs of a cycle. The left side gives $\tau_i$ times the change of the growth rate over each epoch, $\tau_i\,[\Lambda_A(T_A)-\Lambda_A(0)+\Lambda_B(T_B)-\Lambda_B(0)]=2\,\tau_i\,[\Lambda_A(T_A)+\Lambda_B(T_B)-2]$ by Eq.~\eqref{eq:Lambdajump}. Dividing by the cycle duration gives
\begin{equation}
 \overline{\Lambda\,(\Lambda-1)}=\epsilon^2\,(\tau_i-1)\,s_i^2\,\overline{\Var(\psi_i)}-\frac{2\,\tau_i}{T_A+T_B}\,\big[\Lambda_A(T_A)+\Lambda_B(T_B)-2\big],
 \label{eq:cycle}
\end{equation}
where overlines denote averages over a cycle, as in Eq.~\eqref{eq:cycleavg}. This relation is exact, for any offspring distribution and any strength of selection. The first term is the selection term of Eq.~\eqref{eq:Lambdadot}, averaged over the cycle. The second is the cost of switching. In a constant environment, the growth rate does not change on average, and this term vanishes. In a switching environment, each epoch starts from the reversed excess growth rate reached at the end of the previous one, so a response that raised growth at the end of one epoch lowers it at the start of the next.

For weak selection, the variance in Eq.~\eqref{eq:Lambdadot} is one, $\Lambda\,(\Lambda-1)=\Lambda-1$ to leading order, and, for equal epochs, the end values follow from Sec.~\ref{sec:opposite}: by Eq.~\eqref{eq:Lambdasigma} and $\chi_i(T)=\theta_i\,(\tau_i-1)\,s_i$,
\begin{equation}
 \Lambda_A(T)-1=\Lambda_B(T)-1=\epsilon^2\,(\tau_i-1)\,s_i^2\,\theta_i.
 \label{eq:Lambdaend}
\end{equation}
Inserting this into Eq.~\eqref{eq:cycle}, with $\overline{\Lambda\,(\Lambda-1)}=\overline\Lambda-1$ to this order, recovers Eq.~\eqref{eq:opposite} for a single mode.

\section{Optimal inheritance for a single mode}
\label{sec:optimum}
We compare offspring distributions that have the same inheritance modes and differ only in their eigenvalues, so that the selection coefficients do not change as inheritance varies. Gaussian inheritance (Sec.~\ref{sec:gaussian}) is an example, because the Hermite polynomials do not depend on $\rho$. A selection profile proportional to $\psi_i$ acts on mode $i$ alone, by Eq.~\eqref{eq:ortho}, and for opposite profiles in equal epochs Eq.~\eqref{eq:opposite} then becomes
\begin{equation}
 \overline\Lambda=1+\epsilon^2\,s_i^2\,(\tau_i-1)\,F(\tau_i)+O(\epsilon^3),
 \label{eq:singlemode}
\end{equation}
which depends on inheritance only through the relaxation time $\tau_i$ of the selected mode. The product of $\tau_i-1$ and $F$ vanishes when $\tau_i$ is much longer than $T$, because the mode then spends each epoch undoing the response it built in the previous one, and it is negative when $\tau_i$ is shorter than one division time. Growth is therefore fastest at an intermediate value of $\tau_i$, which we denote $\tau_i^*$.

For epochs much longer than a division time, $T\gg1$, the optimum $\tau_i^*$ is also much longer than one division time, so to leading order it maximizes the product of $\tau_i$ and $F$. Setting the derivative of this product with respect to $\tau_i$ to zero, using Eq.~\eqref{eq:F}, gives
\begin{equation}
 \frac{4\,\tau_i}{T}\,\theta_i=2-\theta_i^2.
 \label{eq:stationary}
\end{equation}
We denote the value of $\theta_i$, Eq.~\eqref{eq:thetadef}, at the leading-order optimum by $\theta^*$.
The right-hand side of Eq.~\eqref{eq:stationary} is then $2-(\theta^*)^2$, and solving for $\tau_i$ gives the leading-order optimum
\begin{equation}
 \tau_i^*\approx\frac{2-(\theta^*)^2}{4\,\theta^*}\,T,
 \label{eq:taustar0}
\end{equation}
and inserting it back into Eq.~\eqref{eq:thetadef} shows that $\theta^*$ is the nonzero root of
\begin{equation}
 \theta^*=\tanh\frac{2\,\theta^*}{2-(\theta^*)^2},
 \label{eq:thetastar}
\end{equation}
so that $\theta^*\approx0.9226$. The optimum is therefore proportional to the epoch duration, with slope $[2-(\theta^*)^2]/(4\,\theta^*)\approx0.3113$.

Since $(\tau_i-1)\,F=\tau_i\,F-F$, the remaining term $-F$ shifts the optimum. Linearizing the stationarity condition about Eq.~\eqref{eq:taustar0}, the shift is $F'/(\tau_i\,F)''$, where primes denote derivatives with respect to $\tau_i$ at the leading-order optimum. Because $F$ depends on $\tau_i$ only through $\tau_i/T$, and $\tau_i\,F$ is $T$ times a function of $\tau_i/T$, both $F'$ and $(\tau_i\,F)''$ are proportional to $1/T$, so the shift does not depend on $T$. Simplifying with Eq.~\eqref{eq:thetastar} gives
\begin{equation}
 \tau_i^*=\frac{2-(\theta^*)^2}{4\,\theta^*}\,T+\frac{2-(\theta^*)^2}{2\,(3(\theta^*)^2-2)}+O(T^{-1}).
 \label{eq:taustar}
\end{equation}
Numerically, the offset $[2-(\theta^*)^2]/[2\,(3(\theta^*)^2-2)]\approx1.038$, so the optimal relaxation time is about one division time longer than $0.311\,T$. Direct numerical maximization gives optimal relaxation times of $2.87$, $4.29$, $7.34$, and $32.18$ for epoch durations of $5$, $10$, $20$, and $100$, compared with $2.59$, $4.15$, $7.26$, and $32.17$ from Eq.~\eqref{eq:taustar}. At the optimum, $T/\tau_i^*\approx3.2$, and mode $i$ reaches the fraction $\theta^*\approx0.92$ of its steady response by the end of an epoch. Because $\tau_i^*$ grows with $T$, the weak-selection limit used here also requires $\epsilon^2\,s_i^2\,T$ to be small. Section~\ref{sec:mean} shows that the relevant small parameter is the relative change of the growth rate, which is of order $\epsilon^2\,s_i^2\,\tau_i$.

The optimal growth rate follows by inserting the optimum into Eq.~\eqref{eq:singlemode}. At the leading-order optimum, Eqs.~\eqref{eq:F}, \eqref{eq:thetadef}, and \eqref{eq:taustar0} give the finite-epoch factor
\begin{equation}
 F(\tau_i^*)=1-\frac{2\,\tau_i^*}{T}\,\theta^*=\frac{(\theta^*)^2}{2},
 \label{eq:Fstar}
\end{equation}
so that $F(\tau_i^*)\approx0.43$ and $\tau_i^*\,F(\tau_i^*)=\theta^*\,[2-(\theta^*)^2]\,T/8$. Because the growth rate is stationary at the optimum, the constant shift in Eq.~\eqref{eq:taustar} changes it only at order $1/T$. The optimal growth rate is therefore
\begin{equation}
 \overline\Lambda=1+\epsilon^2\,s_i^2\left[\frac{\theta^*\,\big(2-(\theta^*)^2\big)}{8}\,T-\frac{(\theta^*)^2}{2}\right]+O\big(\epsilon^2\,T^{-1},\epsilon^3\big),
 \label{eq:Lambdaopt}
\end{equation}
where $\theta^*\,[2-(\theta^*)^2]/8\approx0.1325$ and $(\theta^*)^2/2\approx0.43$. Compared with the steady growth rate of Eq.~\eqref{eq:Lambdastar} for a constant environment and the same relaxation time, a switching environment retains the fraction of the selective gain given by Eq.~\eqref{eq:Fstar}, a little less than one half.

Inverting the relaxation time of Eq.~\eqref{eq:hermite} gives the optimal parent--offspring correlation,
\begin{equation}
 (\rho^*)^i=1-\frac{1}{2\,\tau_i^*},
 \label{eq:rhostar}
\end{equation}
which lies between zero and one because the optimal relaxation time exceeds one. It approaches one as epochs lengthen. Because higher modes relax faster at the same $\rho$, the optimal correlation increases with $i$. When selection acts on several modes, their contributions to Eq.~\eqref{eq:opposite} add. Each depends on $\rho$ only through $\tau_i$, which for positive $\rho$ exceeds $1/2$ and increases with $\rho$, and each has a single maximum, at the single-mode optimum. To show this, let $v\equiv T/(2\,\tau_i)$, so that $F(\tau_i)=1-\tanh(v)/v$, which is positive and decreases with $\tau_i$, because $\tanh(v)/v$ decreases with $v$. For $\tau_i<1$, the factor $\tau_i-1$ is negative and increases with $\tau_i$, so the product $(\tau_i-1)\,F(\tau_i)$ increases with $\tau_i$. For $\tau_i>1$, where $v$ lies between zero and $T/2$,
\begin{equation}
 \frac{d\ln\big[(\tau_i-1)\,F(\tau_i)\big]}{d\ln v}=E(v)-\frac{T}{T-2v},\qquad
 E(v)\equiv\frac{\tanh v-v/\cosh^2v}{v-\tanh v}.
 \label{eq:unimodal}
\end{equation}
As $v$ increases from zero to $T/2$, the second term increases from one to infinity, whereas $E$ decreases from two: its derivative has the sign of $2v^2-v\tanh v-\sinh^2v$, which is at most $-2v^6/45$, because $\sinh^2v\geq v^2+v^4/3+2v^6/45$, from its series, and $\tanh v\geq v-v^3/3$, since the difference has the derivative $v^2-\tanh^2v\geq0$. The right side of Eq.~\eqref{eq:unimodal} therefore changes sign once, and the product has a single maximum. The best positive $\rho$ thus lies between the smallest and the largest of the single-mode optima.

Setting the derivative of Eq.~\eqref{eq:opposite} with respect to $\rho$ to zero, with $d\tau_i/d\rho=2\,i\,\rho^{i-1}\,\tau_i^2$ from Eq.~\eqref{eq:hermite}, gives the condition for the best $\rho$,
\begin{equation}
 \sum_{i\geq1} i\,\rho^{i-1}\,\tau_i^2\,s_i^2\,\frac{d\big[(\tau_i-1)\,F(\tau_i)\big]}{d\tau_i}=0.
 \label{eq:rhomulti}
\end{equation}
Differentiating Eq.~\eqref{eq:F}, with $\theta_i$ of Eq.~\eqref{eq:thetadef} and $d\theta_i/d\tau_i=-(1-\theta_i^2)\,T/(2\,\tau_i^2)$, gives $dF/d\tau_i=(1-\theta_i^2)/\tau_i-2\,\theta_i/T$, so that
\begin{equation}
 \frac{d\big[(\tau_i-1)\,F(\tau_i)\big]}{d\tau_i}=2-\theta_i^2-\frac{2\,(2\tau_i-1)\,\theta_i}{T}-\frac{1-\theta_i^2}{\tau_i}.
 \label{eq:dgain}
\end{equation}
Each mode thus pulls $\rho$ toward its own optimum with the weight $i\,\rho^{i-1}\,\tau_i^2\,s_i^2$. For long epochs, $\tau_i\gg1$, the terms of relative order $1/\tau_i$ drop out, and setting Eq.~\eqref{eq:dgain} to zero gives Eq.~\eqref{eq:stationary}.

\section{Other modes of reproduction and selection}
\label{sec:other}

We now allow budding and selection on removal. In budding, a mother keeps her phenotype and produces one daughter. The number of inheritance events per division is
\begin{equation}
 \nu\equiv
 \begin{cases}
 2 & \text{(fission)},\\
 1 & \text{(budding)}.
 \end{cases}
 \label{eq:nu}
\end{equation}
Individuals may also be removed, by death or dilution, at a phenotype-independent rate $d$ without selection. To keep a single selection profile, we let selection on removal lower the removal rate of phenotype $x$ by $\epsilon\,s(x)$, so that in every case a positive selection profile favors a phenotype. Selection that adds directly to the net growth rate has the same form as selection on removal. The expected density obeys
\begin{equation}
\begin{aligned}
 \partial_t n(x,t)&=\int\big[2\,\mathcal P(x|y)-\delta(x-y)\big]\,\big(1+\epsilon\,s(y)\big)\,n(y,t)\,dy-d\,n(x,t) &&\text{(fission)},\\
 \partial_t n(x,t)&=\int \mathcal P(x|y)\,\big(1+\epsilon\,s(y)\big)\,n(y,t)\,dy-d\,n(x,t) &&\text{(budding)},\\
 \partial_t n(x,t)&=\int\big[\nu\,\mathcal P(x|y)-(\nu-1)\,\delta(x-y)\big]\,n(y,t)\,dy-\big(d-\epsilon\,s(x)\big)\,n(x,t) &&\text{(removal)},
\end{aligned}
 \label{eq:other}
\end{equation}
where the first two lines describe selection on the division rate, and the last line describes selection on removal for either mode of reproduction. Budding has no delta function, because the mother survives her division. In all three cases the growth rate is
\begin{equation}
 \Lambda(t)=1-d+\epsilon\int s(x)\,p(x,t)\,dx,
\end{equation}
so the neutral removal rate only lowers the growth rate of every phenotype equally, and the growth-rate expressions of Secs.~\ref{sec:modes} and \ref{sec:switching} hold with their leading term one replaced by $1-d$. A budding mechanism that changes the mother's phenotype would require its own offspring distribution and is not covered here.

Projecting Eq.~\eqref{eq:other} onto the modes as in Sec.~\ref{sec:modeexact} gives exact equations for the mode amplitudes,
\begin{equation}
\begin{aligned}
 \dot \chi_i(t)&=(2\lambda_i-1)\Cov(\psi_i,s)-2\,(1-\lambda_i)\,\langle1+\epsilon\,s\rangle\,\chi_i(t) &&\text{(fission)},\\
 \dot \chi_i(t)&=\lambda_i\Cov(\psi_i,s)-(1-\lambda_i)\,\langle1+\epsilon\,s\rangle\,\chi_i(t) &&\text{(budding)},\\
 \dot \chi_i(t)&=\Cov(\psi_i,s)-\nu\,(1-\lambda_i)\,\chi_i(t) &&\text{(removal)},
\end{aligned}
 \label{eq:chiother}
\end{equation}
where $\langle1+\epsilon\,s\rangle$ is the mean division rate. The neutral removal rate cancels, because it removes all phenotypes equally. Selection on removal does not change how often phenotypes are redrawn, so the relaxation rate of each mode is unchanged by it. In the weak-selection limit, the covariance becomes $s_i$ and the mean division rate one, which gives
\begin{equation}
 \dot \chi_i(t)=q_i-\frac{\chi_i(t)}{\tau_i},\qquad
 \tau_i\equiv\frac{1}{\nu\,(1-\lambda_i)},
 \label{eq:modeother}
\end{equation}
with forcing projections
\begin{equation}
\begin{aligned}
 q_i&\equiv(2\lambda_i-1)\,s_i &&\text{(fission)},\\
 q_i&\equiv\lambda_i\,s_i &&\text{(budding)},\\
 q_i&\equiv s_i &&\text{(removal)}.
\end{aligned}
 \label{eq:qi}
\end{equation}
The steady response is
\begin{equation}
 \chi_i^\infty=\tau_i\,q_i,
 \label{eq:speedresponse}
\end{equation}
which generalizes the speed--response relation of the main text. For symmetric inheritance, the sign of each steady contribution to growth is that of $2\lambda_i-1$ for fission, of $\lambda_i$ for budding, and positive for removal. With Eq.~\eqref{eq:nu}, both modes of reproduction give the same steady response to selection on the division rate,
\begin{equation}
\begin{aligned}
 \chi_i^\infty&=(\tau_i-1)\,s_i &&\text{(fission, budding)},\\
 \chi_i^\infty&=\tau_i\,s_i &&\text{(removal)}.
\end{aligned}
 \label{eq:chiinfother}
\end{equation}
The factor two in the fission forcing matches the two inheritance events per division, so both responses to selection on division are offset by one division time. Selection on removal does not act through inheritance and gives strict proportionality.

The derivation of Sec.~\ref{sec:switching} used only the mode equation and the steady response. In Eq.~\eqref{eq:periodic}, the factor $\tau_i-1$ is therefore replaced by $\tau_i$ for selection on removal, and for opposite profiles in equal epochs
\begin{equation}
\begin{aligned}
 \overline\Lambda&=1-d+\epsilon^2\sum_{i\geq1}(\tau_i-1)\,s_i^2\,F(\tau_i)+O(\epsilon^3) &&\text{(fission, budding)},\\
 \overline\Lambda&=1-d+\epsilon^2\sum_{i\geq1}\tau_i\,s_i^2\,F(\tau_i)+O(\epsilon^3) &&\text{(removal)}.
\end{aligned}
\end{equation}
Selection on division with budding thus depends on the relaxation time in the same way as with fission, and Eq.~\eqref{eq:taustar} gives its optimum. For selection on removal, growth depends on inheritance only through the product of $\tau_i$ and $F$, which is the leading-order problem of Sec.~\ref{sec:optimum}. Its optimum is therefore exact for all epoch durations,
\begin{equation}
 \tau_i^*=\frac{2-(\theta^*)^2}{4\,\theta^*}\,T,
 \label{eq:taustarrem}
\end{equation}
with $\theta^*$ from Eq.~\eqref{eq:thetastar}.
For Gaussian inheritance, inverting Eq.~\eqref{eq:modeother} with $\lambda_i=\rho^i$ gives the optimal parent--offspring correlation,
\begin{equation}
 (\rho^*)^i=1-\frac{1}{\nu\,\tau_i^*}.
 \label{eq:rhostarother}
\end{equation}
For selection on division this lies between zero and one. For selection on removal, the right-hand side must exceed $-1$, and for even $i$ it must be nonnegative. Otherwise, the best accessible correlation lies on the boundary.

\section{Other environments}
\label{sec:env}

\subsection{Environment-dependent mean division rate}
Suppose that the neutral division rate in epoch $B$ is $\Gamma$ times that in epoch $A$, that time is measured in units of the mean division time in epoch $A$, and that selection changes the division rate in relative terms. For fission, Eq.~\eqref{eq:fission} in epoch $B$ is then multiplied by $\Gamma$. Both the relaxation rate and the forcing of each mode are multiplied by this factor, so the steady responses are unchanged, and the growth rate in epoch $B$ is also multiplied by $\Gamma$. Epoch $B$ therefore contributes as an epoch of duration $\Gamma\,T_B$ in its own time units. With a common neutral removal rate, Eq.~\eqref{eq:periodic} becomes
\begin{equation}
 \overline\Lambda=\frac{T_A+\Gamma\,T_B}{T_A+T_B}-d+\frac{\epsilon^2}{T_A+T_B}\sum_{i\geq1}(\tau_i-1)\Big[T_A\,(s_i^A)^2+\Gamma\,T_B\,(s_i^B)^2-L_i\,(s_i^A-s_i^B)^2\Big]+O(\epsilon^3),
 \label{eq:periodicGamma}
\end{equation}
with $T_B$ replaced by $\Gamma\,T_B$ in the switching lag $L_i$. The first two terms are the neutral growth rate averaged over the cycle. Because the offspring distribution is the same in both epochs, so are the inheritance modes.

\subsection{Arbitrary and random environments}
For a selection profile that varies arbitrarily in time, Eq.~\eqref{eq:modeother} has the solution
\begin{equation}
 \chi_i(t)=\int_0^\infty e^{-u/\tau_i}\,q_i(t-u)\,du,
 \label{eq:memory}
\end{equation}
once the initial condition has been forgotten. Each mode thus integrates past selection with a weight that decays exponentially on its relaxation time $\tau_i$. We define the autocorrelation of the selection coefficient of mode $i$,
\begin{equation}
 C_i(u)\equiv\overline{s_i(t)\,s_i(t-u)},
 \label{eq:autocorr}
\end{equation}
where the overline denotes a long-time average, which for a stationary random environment is the stationary expectation. Inserting Eq.~\eqref{eq:memory} into Eq.~\eqref{eq:Lambdamodes} and using Eq.~\eqref{eq:qi}, with a common division rate, gives
\begin{equation}
\begin{aligned}
 \overline\Lambda&=1-d+\epsilon^2\sum_{i\geq1}\frac{\tau_i-1}{\tau_i}\int_0^\infty e^{-u/\tau_i}\,C_i(u)\,du+O(\epsilon^3) &&\text{(fission, budding)},\\
 \overline\Lambda&=1-d+\epsilon^2\sum_{i\geq1}\int_0^\infty e^{-u/\tau_i}\,C_i(u)\,du+O(\epsilon^3) &&\text{(removal)},
\end{aligned}
 \label{eq:Lambdageneral}
\end{equation}
where the overline on $\Lambda$ also denotes the long-time average. The prefactor in the first line is the forcing coefficient of Eq.~\eqref{eq:qi}, written in terms of the relaxation time. It is the same for fission and budding. A constant environment recovers Eq.~\eqref{eq:Lambdastar}. Opposite profiles in equal epochs have an autocorrelation that is a triangle wave, and Eq.~\eqref{eq:Lambdageneral} recovers Eq.~\eqref{eq:opposite}.

Equation~\eqref{eq:Lambdageneral} shows when an intermediate strength of inheritance can be optimal. If the autocorrelation is nonnegative at all lags, the integral is positive and increases with the relaxation time, and so does the prefactor, which for fission and budding changes sign at $\tau_i=1$. The contribution of each mode is then increasing for $\tau_i>1$ and, for fission and budding, negative for $\tau_i<1$, so there is no interior optimum at this order. This does not show that the slowest inheritance is best, because the fluctuations of slow modes grow with $\tau_i$ until the weak-selection expansion fails. An interior optimum, as in Sec.~\ref{sec:optimum}, requires selection that is anticorrelated at some lag, as it is when the environment reverses selection at regular intervals.

\subsection{Markov environments}
Let the environment be a Markov chain with states $e$, independent of the population, as in models of stochastic phenotype switching~\cite{Kussell2005,Skanata2016}, with rates $Q_{eh}$ from state $h$ to state $e$ for distinct states, and stationary probabilities $\pi_e$,
\begin{equation}
 \sum_e Q_{eh}=0,\qquad \sum_h Q_{eh}\,\pi_h=0,\qquad \sum_e\pi_e=1.
 \label{eq:markov}
\end{equation}
Let $s_i^{(e)}$ be the selection coefficient of mode $i$ in state $e$, with a common division rate in all states. The stationary autocorrelation is
\begin{equation}
 C_i(u)=\sum_{e,h}s_i^{(e)}\,\big[e^{Q\,u}\big]_{eh}\,\pi_h\,s_i^{(h)},
\end{equation}
and the integral in Eq.~\eqref{eq:Lambdageneral} becomes a resolvent,
\begin{equation}
 \int_0^\infty e^{-u/\tau_i}\,C_i(u)\,du=\sum_{e,h}s_i^{(e)}\,\Big[\big(I/\tau_i-Q\big)^{-1}\Big]_{eh}\,\pi_h\,s_i^{(h)}.
 \label{eq:resolvent}
\end{equation}
With environment-dependent division rates, the relaxation rate and forcing of each mode are multiplied by the division rate of the current state, and the averages of the mode amplitudes jointly with the state solve a linear system of the same form.

As an example, consider two states with opposite selection profiles that switch at the same rate in both directions, so that each epoch lasts an exponentially distributed time with mean $\overline T$,
\begin{equation}
 Q=\frac{1}{\overline T}\begin{pmatrix}-1&1\\1&-1\end{pmatrix}.
\end{equation}
The autocorrelation decays as an exponential with rate twice the switching rate, and Eq.~\eqref{eq:Lambdageneral} gives, for selection on the division rate,
\begin{equation}
 \overline\Lambda=1-d+\epsilon^2\sum_{i\geq1}\frac{\tau_i-1}{1+2\,\tau_i/\overline T}\,s_i^2+O(\epsilon^3).
 \label{eq:markov2}
\end{equation}
Compared with periodic switching, Eq.~\eqref{eq:opposite}, the finite-epoch factor $F$ is replaced by a factor that decays only inversely with the relaxation time. Each contribution increases monotonically with the relaxation time and, for slow modes, saturates at a value set by half the mean epoch duration. Random switching with exponentially distributed epochs therefore does not favor an intermediate relaxation time at this order, whereas periodic switching does, in line with the criterion above.

\section{Numerical methods}
\label{sec:numerics}

The figures of the main text compare the theory with the exact expected-density equation for Gaussian inheritance, Eq.~\eqref{eq:gausskernel}, and binary fission, without selection [Fig.~1(b)] or with directional selection, $s(x)=x$, or its reverse. The code and data are available at \url{https://github.com/hhmattingly/inheritance_tradeoff}. We discretize the phenotype and normalize the offspring distribution on the grid, so that inheritance conserves the population. The offspring distribution narrows as the parent--offspring correlation approaches one, so for slow modes we refine the grid until the results no longer change. Equation~\eqref{eq:fission} then becomes a linear system for the vector of densities at the grid points,
\begin{equation}
 \dot n(t)=\mathsf A\,n(t),\qquad
 \mathsf A_{x'x}\equiv\big[2\,\mathcal P(x'|x)-\delta_{x'x}\big]\,\big(1+\epsilon\,s(x)\big),
\end{equation}
where $\delta_{x'x}$ is the Kronecker delta, and the division rate is set to zero where it would be negative. Its solution, obtained with matrix exponentials, contains all orders in $\epsilon$. The steady state under constant selection is the eigenvector of $\mathsf A$ with the largest eigenvalue. For selection that reverses every $T$, with generators $\mathsf A_A$ for $s(x)=x$ and $\mathsf A_B$ for $s(x)=-x$, the long-time growth rate is
\begin{equation}
 \overline\Lambda=\frac{\ln\mu}{2T},
\end{equation}
where $\mu$ is the largest eigenvalue of the propagator over one cycle, $e^{\mathsf A_B T}e^{\mathsf A_A T}$. Its eigenvector is the periodic state, whose mean division rate gives the growth rate over the cycle in Fig.~3(a) of the main text.

\bibliography{references}